\documentclass[conference]{IEEEtran}

\IEEEoverridecommandlockouts
\usepackage{verbatim}
\usepackage{longtable}
\DeclareMathAlphabet{\mathpzc}{OT1}{pzc}{m}{it}
\usepackage{amsmath}
\usepackage{amsfonts}
\usepackage{amssymb}
\usepackage{multirow}
\usepackage{amsmath,amssymb,amsfonts}
\usepackage{fancyhdr}
\usepackage[ruled,vlined,linesnumbered]{algorithm2e}
\usepackage[
top    = 0.7in,
bottom = 0.95in,
left   = 0.65in,
right  = 0.59in]{geometry}
\usepackage{graphicx}
\usepackage{textcomp}
\usepackage{xcolor}
\usepackage{subcaption}
\usepackage{algpseudocode}
\usepackage{enumitem}
\usepackage{float}
\restylefloat{figure}
\usepackage{graphicx}
\usepackage[normalem]{ulem}
\useunder{\uline}{\ul}{}

\begin{document}

\title{An Efficient and Scalable Auditing Scheme for Cloud Data Storage using an Enhanced B-tree}

\author{\IEEEauthorblockN{Tariqul Islam$^{1}$, Faisal Haque Bappy$^{2}$,  Md Nafis Ul Haque Shifat$^{3}$, Farhan Ahmad$^{4}$, \\ Kamrul Hasan$^{5}$, and Tarannum Shaila Zaman$^{6}$}
\IEEEauthorblockA{
$^{1, 2}$ Syracuse University, Syracuse, NY, USA\\
$ ^{3}$ Massachusetts Institute of Technology, MA, USA\\
$ ^{4}$ New Gov. Degree College, Rajshahi, Bangladesh\\
$ ^{5}$ Tennessee State University, Nashville, TN, USA\\
$ ^{6}$ SUNY Polytechnic Institute, NY, USA\\
Email: \{mtislam, fbappy\}@syr.edu,  \{nafis.shifat1, farhan.ru.school\}@gmail.com, \\ and \{mhasan1@tnstate, zamant@sunypoly\}.edu} 
}


\maketitle

\thispagestyle{fancy}
\lhead{This work has been accepted at the 2024 IEEE International Conference on Communications (ICC)}
\cfoot{}

\begin{abstract}
An efficient, scalable, and provably secure dynamic auditing scheme is highly desirable in the cloud storage environment for verifying the integrity of the outsourced data. Most of the existing work on remote integrity checking focuses on static archival data and therefore cannot be applied to cases where dynamic data updates are more common. Additionally, existing auditing schemes suffer from performance bottlenecks and scalability issues. To address these issues, in this paper, we present a novel dynamic auditing scheme for centralized cloud environments leveraging an enhanced version of the B-tree. Our proposed scheme achieves the immutable characteristic of a decentralized system (i.e., blockchain technology) while effectively addressing the synchronization and performance challenges of such systems. Unlike other static auditing schemes, our scheme supports dynamic insert, update, and delete operations. Also, by leveraging an enhanced B-tree, our scheme maintains a balanced tree after any alteration to a certain file, improving performance significantly. Experimental results show that our scheme outperforms both traditional Merkle Hash Tree-based centralized auditing and decentralized blockchain-based auditing schemes in terms of block modifications (e.g., insert, delete, update), block retrieval, and data verification time.
\end{abstract}

\begin{IEEEkeywords}
Cloud Auditing, Enhanced B-tree, Merkle Hash Tree, Blockchain, Persistency, Immutability
\end{IEEEkeywords}

\section{Introduction}

Over the past two decades researchers have been working on solving the data integrity verification issue by proposing various auditing algorithms that can be classified mainly into two categories, namely, i) static model, and ii) dynamic model. Static models \cite{Bowers-2008,Ateniese-2007,shacham2013compact,Ateniese-2008} can perform auditing only on the static archival data which is a serious drawback since data are frequently updated and modified in the cloud. To overcome this issue, several dynamic auditing schemes \cite{Wang-2009-3,Wang-2011,Erway-2009,Yang-2013} are proposed. However, some of them \cite{Wang-2009-3,Wang-2011} are not privacy-preserving and leak sensitive client data to the auditor. Some schemes \cite{Erway-2009,Yang-2013} incur high computation costs and storage overhead on the server side. The B-tree also offers an efficient and dynamic data structure extensively employed in storage disks and database systems \cite{btree-core, btreemdp}. Nevertheless, its application in cloud auditing has been restricted due to some space limitations which this paper addressed. 

Data integrity verification usually requires a considerable amount of resources for computation and communication. Thus, a third-party auditor (TPA) is typically delegated by a client to perform the verification on behalf of data owners, which helps to reduce the overhead in computation, communication, and storage resources at the client side \cite{Shen-2018}. Most of the recently proposed public auditing schemes \cite{Shen-2019,zhang2018enabling,wu2019cpda,xu2020intrusion,gudeme2021certificateless,Shen-2018,fu2017npp} employ a TPA to do the integrity checking. However, the presence of a TPA also brings new security risks, because the TPA can collect information on the outsourced data during the auditing process. Therefore, the TPA cannot be fully trusted. As such it is also essential to guarantee data privacy against the TPA, because the users may store sensitive or confidential files in the cloud.


In recent times, blockchain-based decentralized auditing schemes have been studied because of their transparent and immutable nature. These solutions can certify proof of existence and detect unauthorized alterations of cloud data \cite{fran}. 
The cost of data verification in such decentralized schemes becomes expensive because of the time-sensitive communication, coordination, and synchronization among the peer nodes \cite{garg}. Despite offering immutability and transparency in a decentralized way, blockchain-based data verification schemes suffer from computation overhead and performance bottlenecks \cite{deep}. Therefore, if we can incorporate the immutability and transparency features of blockchain into a centralized cloud environment, we could eliminate the synchronization and performance issues while offering a secure, tamper-proof, and auditable cloud storage scheme.

It is quite evident from the above discussions that there is no ``one-size-fits-all” solution. This is an open problem and there is still room to contribute. In this work, our goal is to design and develop a cloud auditing scheme that can address most of the issues mentioned above. To carry out our research agenda, in this paper, we propose a centralized dynamic auditing scheme using an enhanced B-tree and we named it \texttt{EB-tree}. It can facilitate file version control, dynamic data update operations (e.g., insert, delete, modification), and efficient batch auditing of user data with the assistance of a semi-trusted third-party auditor. The computationally intensive auditing tasks are delegated to semi-trusted TPA, requiring only lightweight computation at the client end to verify data integrity. TPA conducts the auditing process with client-provided minimal file metadata, prohibiting TPA's access to the original file (or blocks) owned by clients and thus, ensuring data confidentiality and privacy. 

The following are the major contributions of our work:
\begin{itemize}

\item \textbf{Data Dynamics:} We proposed a novel storage auditing scheme \texttt{EB-tree} utilizing an enhanced B-tree to enable version control, persistency, and dynamic auditing in a centralized cloud environment.
\item \textbf{Enhanced Throughput:} Our \texttt{EB-tree} maintains a balanced tree after each insert/update/delete operation; thus enhancing dynamic data modification speed compared to traditional auditing schemes.
\item \textbf{Scalability:} We developed a prototype of \texttt{EB-tree}, and compared batch auditing performance with the existing schemes. \texttt{EB-tree} significantly outperforms all schemes and can produce auditing results in less than a second.  
\item \textbf{Security and Integrity:} \texttt{EB-tree} ensures data security and integrity through the use of cryptographic hashes and randomized seed-based batch auditing.
\end{itemize}

The remainder of the paper is organized in the following order. The status of our work compared to the literature is presented in section II. The overview of the proposed architecture is described in section III. In section IV, we present our implementation overview and then the system evaluation is presented in section V. Finally, in section VI, we conclude the paper.

\section{Related Works}
\textbf{Static Auditing (On Archival Data).} Zhu et al. \cite{zhu} propose a signature-based architecture for cloud data integrity verification, but it relies on random masking techniques and lacks support for auditing in multi-replica cloud environments. Zeng ~\cite{Zeng-2008} presents a provable data integrity (PDI) scheme that restricts incremental fingerprinting and prevents data file modification once fingerprinted. Ateniese et al.'s Scalable Provable Data Possession (SPDP) technique supports certain operations but has limitations on updates and challenges, restricting its practicality for large files. Erway et al. ~\cite{Erway-2009} introduces the Dynamic Provable Data Possession (DPDP) scheme, but it incurs heavy computation overhead on the client side, raising feasibility concerns in practical data storage implementations.

\textbf{Dynamic Auditing (In case of Frequent Data Updates).} Garg et al. \cite{garg} introduce a MHT (Merkle Hash Tree)-based public auditing scheme using third-party auditors (TPA). However, the trustworthiness assumption of TPAs raises security concerns. Our proposed architecture addresses this by minimizing user-provided metadata and enhancing information privacy. Wang et al. ~\cite{Wang-2009-3, Wang-2011} explore dynamic data modification using a classic Merkle Hash Tree ~\cite{Tilborg-2011} construction but risk data corruption by the auditor. B-tree serves as a dynamic data structure facilitating efficient dynamic operations \cite{btree-core} and is commonly employed for enhanced storage performance in both magnetic and SSD (Solid-State Drive) environments \cite{ssdbtree}. Nonetheless, this paper identifies and discusses certain space limitations associated with the B-tree and proposes an enhanced version of it for cloud auditing.

\textbf{Blockchain-based Auditing.} Because of the improved security, traceability, and immutability, blockchain-based solutions are becoming widely adopted for cloud auditing environments \cite{shaf,jiax,ped,dan}. Various approaches, including encrypted storage, off-chain data storage with on-chain hashes, and role-based access control, are explored\cite{fran}. A consortium blockchain solution based on zero-knowledge proofs is proposed in the literature to improve auditability \cite{xu}. Zhang et al. \cite{zhang-bc-2022} propose a smart-contract-based auditing scheme without a TPA but with storage overheads. Francati et al.'s \cite{dan} off-chain storage-based auditing addresses rational behavior assumptions but requires added security for data transfer and faces network latency challenges.

\section{Architectural Overview} Our proposed storage auditing scheme leverages the enhanced \texttt{B-tree} data structure. In this section, we first present some preliminaries on the conventional \texttt{B-tree} and then demonstrate its limitations in practical use cases. Then we provide an overview of our approach. 
\subsection{B-tree}
A B-tree is a self-balancing tree data structure that generalizes binary search trees by accommodating multiple keys in a single node. It allows basic operations like update, insertion, and deletion in an efficient manner, specifically in logarithmic time \cite{cormen2022introduction}. Each node, excluding leaves, has the following properties, i) number of keys in a node ($n$), ii) keys are stored in non-decreasing order so that $k_1 <= k_2 … <= k_n$ must be true, and iii) a boolean value indicates whether the node is a leaf or an internal node. Additionally, each node, except the leaf nodes, has $(n + 1)$ children $C_1, C_2, … C_{n+1}$. The keys in the sub-tree of children $C_i$ will be between $k_i$ and $k_{i + 1}$, dividing the keys in specific ranges which allows finding keys efficiently. 

The tree adheres to strict bounds dictated by a minimum degree $t$, ensuring that nodes, except the root, have at least $t - 1$ keys and at least $t$ children. With a cap of $2t - 1$ keys per node, and $2t$ as the maximum number of children, B-trees maintain balance and efficiency in operations like insertion, deletion, and updates \cite{btree-core}. It is widely used in file and database systems (e.g., ReiserFS, XThere, MySQL) due to its easy construction and efficient retrieval \cite{btreemdp}. It maintains tree balance by heavily overwriting data in the same node and is also effective for SSDs \cite{ssdbtree}. In cloud auditing, handling both archival and non-archival data using SSDs and regular magnetic disks is crucial. Hence, we opt for the B-tree like data structure over alternatives like MHT for its versatility and proven effectiveness in various storage scenarios.

\begin{figure}[ht]
\centering
\includegraphics[width=1.8in, height=0.7in]{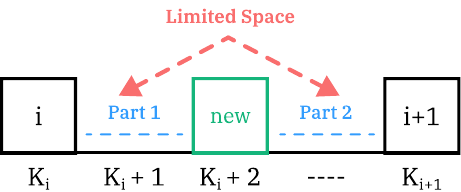}
        \caption{Space Limitation in conventional B-tree} \label{fig:issue}
\end{figure}

\subsection{Limitations of Conventional B-tree}
In a conventional B-tree structure, data can be stored as a pair of keys and data blocks. But this raises a significant limitation: \textit{if there already exist two blocks in the tree with consecutive key values, say $x$ and $(x + 1)$, we cannot insert a new block between them.} If $K_i$ is the key of the $i^{th}$ block, and we want to insert a block between $i$ and $(i+1)^{th}$ blocks, we must choose a key within the range $(k_i, k_{i+1})$. That means we have a limited number of keys left to use to insert blocks between $i$ and $(i+1)^{th}$ block. In fact, there can be a scenario where only after $log_2(S)$ insertions, there will be two blocks with consecutive key values, making it impossible to insert any blocks between them. Here $S$ is the number of available keys between $K_i$ and $K_{i+1}$, which is $K_{i+1} - K_i - 1$. Figure \ref{fig:issue} shows a visual representation of this scenario. Here after each insertion, we are dividing the space between them into two parts. So, after each insertion, the smaller partition will have strictly smaller than $(S / 2)$ keys available to use.  If we insert another block in the smaller half, we will get another part that has a space of $(S / 4)$ and so on. Hence, after $log_2(S)$ insert operations, there will be no available key to use. 

\begin{figure}[ht]
\centering
\includegraphics[width=3.1in,height=1.2in]{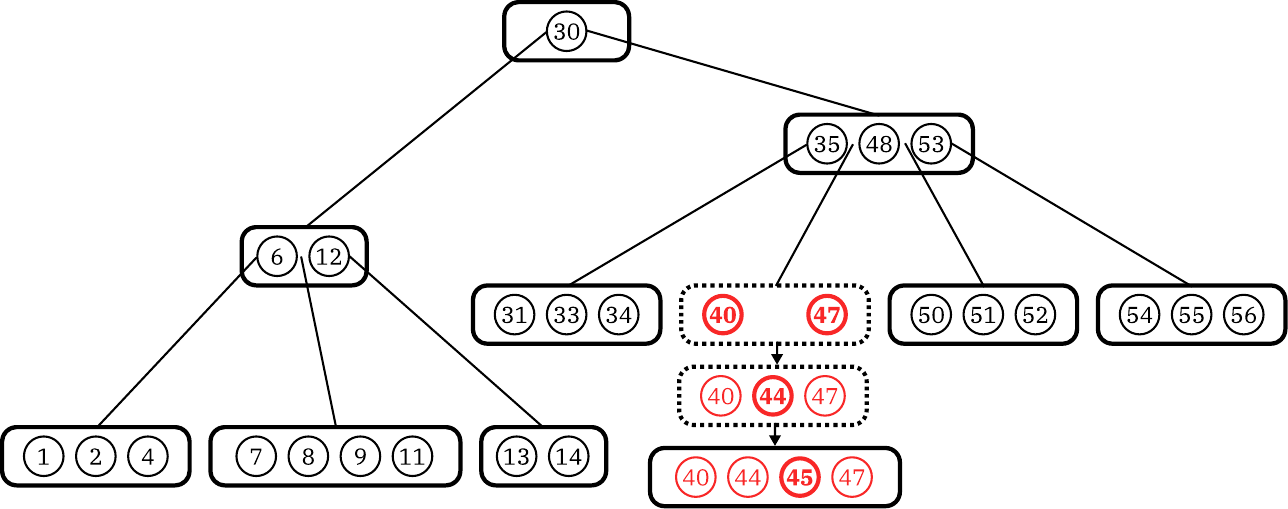}
        \caption{Demonstration of the space limitation} \label{fig:insert}
\end{figure}

\textbf{Demonstration of the scenario.} Initially we are considering a regular B-tree like Figure \ref{fig:insert}. Each key in the tree points to a block (not shown in Figure \ref{fig:insert}). Now if we want to insert a block between the blocks with keys $40$ and $47$, we must choose a key $K$ in the range $[41, 46]$. To make the smaller partition as large as possible, let's assume we chose $K = 44$. Now if we want to insert another block between the blocks with keys $44$ and $47$, we have only two possibilities left $45$ or $46$. If we choose $45$, then clearly we can no longer insert any block between the blocks with keys $44$ and $45$. If we chose $46$ earlier, the same would happen with the keys $46$ and $47$. So, within only $2$ operations, we have reached a point where two keys with consecutive values exist in the tree and we no longer can insert any blocks between them. Note that initially we had $(47 - 40 - 1 = 6)$ usable keys, and since $log_2(6) = 2$, we reached that point only in $2$ operations.

\subsection{Our Enhanced B-tree Approach}
Our proposed \texttt{EB-tree} effectively addresses the limitation of conventional B-tree, enabling the insertion of blocks at any position in the tree while upholding the fundamental characteristics of a B-tree. In this enhanced version, we will directly insert data blocks without explicitly assigning any key to them. Each node except the root must contain at least $(t – 1)$ blocks, and each node can contain at most $(2t – 1)$ blocks. Blocks inside a node will also be sorted in order. For example, if the node has $n$ blocks $B_1, B_2, … , B_n$, then in the actual order of the blocks, $B_i$ will come before $B_{i+1}$ for all $i$s. Similar to the conventional approach, the node (except the leaf nodes) will have $(n + 1)$ children $C_1, C_2, …, C_{n + 1}$, and blocks in $C_i$ will come before $B_i$ in the actual order. This allows us to uniquely determine the order of the blocks without explicitly using any keys. The order will be, $C_1.blocks\rightarrow B_1\rightarrow C_2.blocks\rightarrow B_2\rightarrow …\rightarrow B_n\rightarrow C_{n+1}.blocks$.

To allow insert/update/delete in a certain position, we additionally need to store the number of blocks in each sub-tree of the tree. If the number of blocks in the sub-tree of $C_1, C_2, …, C_{n+1}$ is $S_1, S_2, …, S_{n+1}$ respectively, and we are looking for the $k^{th}$ block, we can easily determine which child of the node will contain that block or if that block is inside the current node. This allows us to efficiently find out the $i^{th}$ block of the tree, delete or update it, or add a new block after it without changing the time complexity of any operations in the traditional B-tree.

\IncMargin{1em}
\setlength{\textfloatsep}{0pt}
 \begin{algorithm}
  \caption{\textbf{\texttt{insertBlock()}}} 
  \label{algo:insert}
  \SetKwFunction{BuildTree}{insertBlock}
  \SetKwInOut{Input}{Input}
  \SetKwInOut{Output}{Output}
  \Indm 
    \Input{The Current Node ($node$), Position of the Block ($p$), The new Block ($B$)} 
    \Output{Inserts the new block ($B$) in the given position ($p$)}
    \Indp
    $n \leftarrow \texttt{totalBlocks}(node)$\\
    
    \eIf{\texttt{isLeaf}($node$)} {
        $i \leftarrow n - 1$\\
        \While {$i >= p$}{
            $node.block[i+1] \leftarrow node.block[i]$\\
            $i \leftarrow i-1$\\
        }
        $node.block[i+1] \leftarrow B$\\
        $\texttt{updateAttributes}(node)$\\
        return\\
    }{
        $id \leftarrow \texttt{getChildId}(node, p)$\\
        \If{$\texttt{isFull}(node.child[id]$} {
            $\texttt{split}(node.child[id])$
        }
        
        \For{$i = 0$ to $(id - 1)$}{
            $size \leftarrow \texttt{subTreeSize}(node.child[i])$\\
            $B_{left} \leftarrow B_{left} + size + 1$\\
        }
        
        $\texttt{insertBlock}(node.child[id], p - B_{left}, B)$\\
        $\texttt{updateAttributes}(node)$\\
    }
\end{algorithm}
\DecMargin{1em}

\subsection{System Components}
\textbf{Client.} Users encrypt files with a private key before uploading them to the cloud. The client informs a Third Party Auditor (TPA) about file metadata for confirmation and updates the TPA on any modifications.

\textbf{Third Party Auditor.} TPAs validate file existence in the cloud on behalf of clients, using challenge messages based on metadata. They verify encrypted user data integrity and conduct periodic audits using client-provided metadata for file modifications.

\textbf{Server.} The server stores encrypted files, ensuring integrity and consistency. It uses an Enhanced B-tree architecture for file block maintenance. When challenged by TPAs, the server generates proof messages to confirm file existence and integrity.

\section{Implementation Overview} In this section, we describe the four fundamental operations that are the core of our auditing scheme. 

\IncMargin{1em}
\setlength{\textfloatsep}{0pt}
 \begin{algorithm}
  \caption{\textbf{\texttt{deleteBlock()}}} 
  \label{algo:delete}
  \SetKwFunction{BuildTree}{deleteBlock}
  \SetKwInOut{Input}{Input}
  \SetKwInOut{Output}{Output}
  \Indm 
    \Input{The Current Node ($node$), Position of the Block to be deleted($p$)} 
    \Output{Deletes the block at the given position ($p$)}
    \Indp
    
    $n \leftarrow \texttt{totalBlocks}(node)$\\
    $id \leftarrow \texttt{getChildId}(node, p)$\\
    
    \For{$i = 0$ to $(id - 1)$}{
        $size \leftarrow \texttt{subTreeSize}(node.child[i])$\\
        $B_{left} \leftarrow B_{left} + size + 1$\\
    }

    $p_{c} \leftarrow B_{left} + \texttt{subTreeSize}(node.child[id]) +1$\\

    \eIf{$id < n$ \&\& $p_{c} == p$} {
        $\texttt{delete}(id)$\\
        return\\
    }{
        $n_{id} \leftarrow \texttt{totalBlocks}(node.child[id])$\\
        \If{$n_{id} == t-1$}{ \Comment{\textcolor{gray}{$t$ = the minimum degree of a node}}\\ 
            $\texttt{fillChild}(node.child[id])$\\
            $\texttt{updateAttributes}(node)$\\
        }

        \If{$id > n$} { 
            $id \leftarrow id - 1$\\
        }
    
        \For{$i = 0$ to $(id - 1)$}{
            $size \leftarrow \texttt{subTreeSize}(node.child[i])$\\
            $B_{left} \leftarrow B_{left} + size + 1$\\
        }

        $\texttt{deleteBlock}(node.child[id], p-B_{left})$\\
        $\texttt{updateAttributes}(node)$\\
    }
    
\end{algorithm}
\DecMargin{1em}
\textbf{Insertion [Algorithm-1].} According to the B-tree properties, a new block is always inserted in a leaf node. For inserting a new block, we efficiently look for the leaf that should contain the new block so that its actual order is preserved. First, we first check if the current node is a leaf or not (line 2). If it is a leaf node, we calculate the number of blocks we are leaving before with a loop to determine the new position to insert the block (lines 3-6). Then we insert the new block in that position and call a function named \texttt{updateAttributes()} (line 8), which updates the sub-tree size, new hashes, and the number of blocks. If the current node is not a leaf, we split the children of the node (lines 11-13) and then calculate the number of blocks we are leaving before with a loop (lines 14-17) to determine the new position. Then we recursively call the \texttt{insertBlock()} function with the new position (Line 19). The complexity of insertion is $O(t*log_t(N))$.

\textbf{Deletion [Algorithm-2].} Similar to insertion, we use a loop to look for the node where the block is currently located (lines 3-6). However, unlike insertion, we need to deal with both leaf and non-leaf nodes for deletion. If the node is a leaf, we immediately delete the block (line 9). Otherwise, we first check for the children of the current node (lines 12-16) and determine the position of the blocks of the child nodes (lines 19-22), and recursively call the \texttt{deleteBlock()} function (line 23) until it finds the leaf node.  

\textbf{Update [Algorithm-3].} The update function is almost similar to the insertion and deletion. Here we also use a loop to look for the node where the block is currently located (lines 3-6). Then if the node is a leaf node, we replace the current block with the new block (line 9). Otherwise, we recursively call the \texttt{updateBlock()} function for the child of the node and the determined position (line 12).

\textbf{Auditing [Algorithm-4].} For auditing, we utilized the concept of sibling path to calculate the root hash, which the user can check to determine the integrity of a data block. If the user asks to audit the block at position $i$, the $i^{th}$ data block and the sibling path are returned. Then the user calculates the hash of the data block, and finally using the sibling path, the user can calculate the root hash of the B-tree and check with his previously stored root hash to determine the integrity of data. In Algorithm \ref{algo:audit}, we calculate two hash values, prefix hash ($H_{pre}$)  and suffix hash ($H_{suf}$) by traversing the sibling paths of the current node (lines 6-16). Then we check if the block is in a leaf node. If it is in a leaf node, we store the calculated prefix and suffix hashes in the $siblingPath$ stack and return the block to the user (lines 18-21). Otherwise, we store the suffix and prefix hashes in the $siblingPath$ stack and call the \texttt{audit()} function recursively for the child of the node (lines 23-26).   

Due to the dynamic nature of the tree, the root hash changes with each operation, necessitating users to update their stored root hash accordingly. This dynamic behavior poses a risk of unauthorized changes by the cloud provider before a new operation, potentially undetected by the user. To address this, users use a private seed before hashing a block, preventing the provider from calculating the hash without knowledge of the seed. This ensures that any unauthorized modification results in a mismatch between the calculated root hash and the stored one, providing a safeguard against undetected alterations.

\IncMargin{1em}
\setlength{\textfloatsep}{0pt}
 \begin{algorithm}
  \caption{\textbf{\texttt{updateBlock()}}} 
  \label{algo:update}
  \SetKwFunction{BuildTree}{updateBlock}
  \SetKwInOut{Input}{Input}
  \SetKwInOut{Output}{Output}
  \Indm 
    \Input{The Current Node ($node$), Position of the Block ($p$), The new Block ($B$)} 
    \Output{Updates the existing block at position ($p$) with the new block ($B$)}
    \Indp
    
    $n \leftarrow \texttt{totalBlocks}(node)$\\
    $id \leftarrow \texttt{getChildId}(node, p)$\\
    
    \For{$i = 0$ to $(id - 1)$}{
        $size \leftarrow \texttt{subTreeSize}(node.child[i])$\\
        $B_{left} \leftarrow B_{left} + size + 1$\\
    }

    $p_{c} \leftarrow B_{left} + \texttt{subTreeSize}(node.child[id]) +1$\\

    \eIf{$id < n$ \&\& $p_{c} == p$} {
        $node.block[id] \leftarrow B$\\
        return\\
    }{
        $\texttt{updateBlock}(node.child[id], p-B_{left}, B)$\\
        $\texttt{updateAttributes}(node)$\\
    }
    
\end{algorithm}
\DecMargin{1em}

\section{System Evaluation}

\subsection{Experimental Setup}
\begin{figure}[t]
\centering
\includegraphics[width=2.in,height=1.3in]{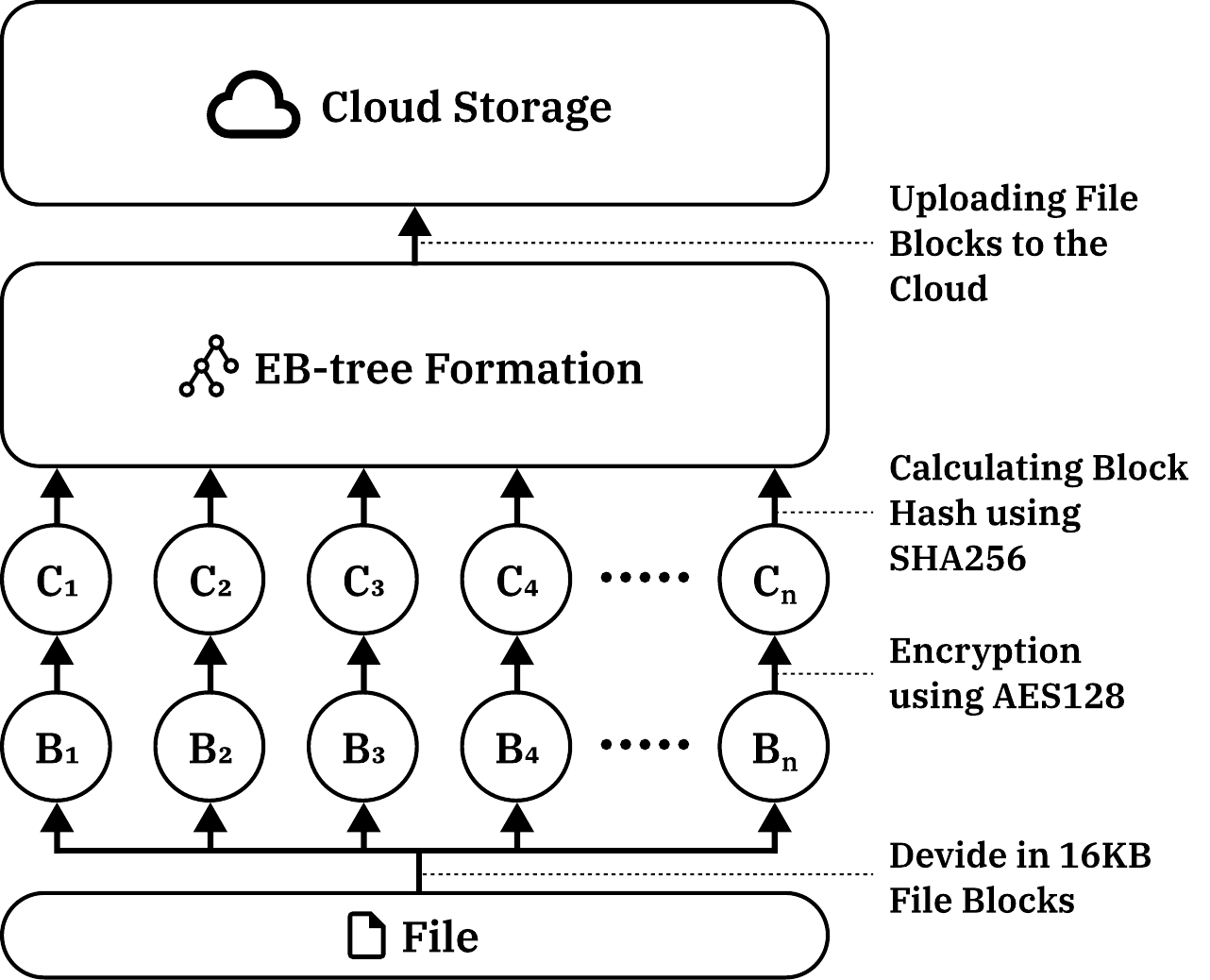}
\caption{System Architecture}  \label{fig:arch}
\end{figure}

We evaluated our dynamic auditing framework in an experimental cloud environment. The system architecture is shown in Figure \ref{fig:arch}. Files were divided into 16 KB blocks and encrypted with AES (Advanced Encryption Standard) using a user-generated 32-byte key. The Third Party Auditor (TPA) received hash values and root node hash metadata for file modifications. We used SHA-256 for collision-resistant hashing. Performance was assessed on an Ubuntu 22.04 machine with an AMD Ryzen-5 5600X 3.7GHz Processor and 32GB of RAM. To facilitate a fair comparison, we implemented, (i) \textit{Conventional MHT}: that uses a tree structure to track hashes for individual data blocks(\cite{liu2023dynamic,gladston2020merkle}), (ii) \textit{8MHT}: a variation of Merkle Hash Tree with 8 branching nodes\cite{YUE20201}, and (iii) \textit{Blockchain with Off-chain Storage}: that utilizes Hyperledger Fabric for auditing large cloud files with off-chain storage in an FTP server\cite{dan}.

\IncMargin{1em}
\setlength{\textfloatsep}{0pt}
 \begin{algorithm}
  \caption{\textbf{\texttt{audit()}}} 
  \label{algo:audit}
  \SetKwFunction{BuildTree}{audit}
  \SetKwInOut{Input}{Input}
  \SetKwInOut{Output}{Output}
  \Indm 
    \Input{The Current Node ($node$), Position of the Block ($p$)} 
    \Output{Returns the Block with the Sibling Path}
    \Indp

    $n \leftarrow \texttt{totalBlocks}(node)$\\
    $id \leftarrow \texttt{getChildId}(node, p)$\\

    \For{$i = 0$ to $(id - 1)$}{
        $H_c \leftarrow \texttt{hash}(node.child[i])$\\
        $H_b \leftarrow \texttt{hash}(node.block[i])$\\
        $H_{pre} \leftarrow H_{pre} + H_c + H_b$\\
        $size \leftarrow \texttt{subTreeSize}(node.child[i])$\\
        $B_{left} \leftarrow B_{left} + size + 1$\\
    }

    \For{$i = (id+1)$ to $(\texttt{totalBlocks}(node)-1)$}{
        $H_c \leftarrow \texttt{hash}(node.child[i])$\\
        $H_b \leftarrow \texttt{hash}(node.block[i])$\\
        $H_{suf} \leftarrow H_{suf} + H_c + H_b$\\
    }

    $H_{suf} \leftarrow H_{suf} + node.child[n]$\\
    $p_{c} \leftarrow B_{left} + \texttt{subTreeSize}(node.child[id]) +1$\\
    
    \eIf{$id < n$ \&\& $p_{c} == p$}{
        $H_{pre} \leftarrow H_{pre} + \texttt{hash}(node.child[id])$\\
        $siblingPath.\texttt{push}(\{H_{pre}, H_{suf}\})$\\
        return $node.block[id]$\\
    }{
        \If{$id < n$}{
            $H_{suf} \leftarrow H_{suf} + \texttt{hash}(node.child[id])$\\
        }
        $siblingPath.\texttt{push}(\{H_{pre}, H_{suf}\})$\\
        return $\texttt{audit}(node.child[id], p-B_{left})$\\
    }
    
\end{algorithm}
\DecMargin{1em}

\begin{figure*}[ht]
\centering
\includegraphics[width=7in, height=1.6in]{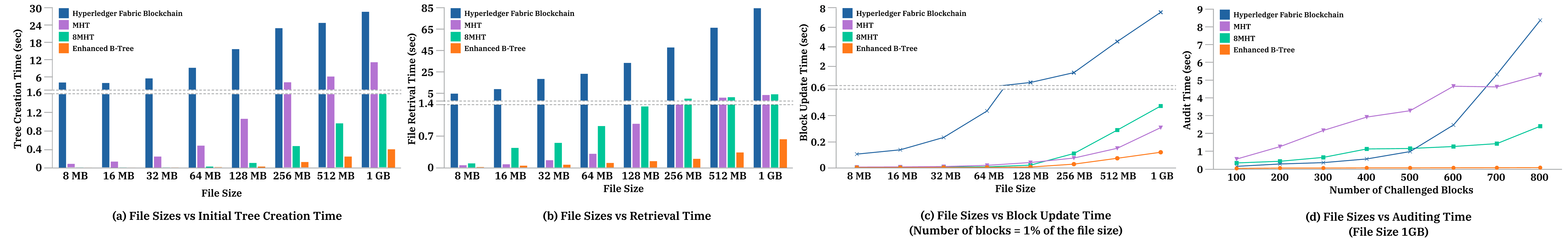}
        \caption{Performance Analysis of EB-tree} \label{fig:perf}
\end{figure*}

\subsection{Performance Analysis}
The evaluation considered four metrics: initial tree creation time, file retrieval time, block update time, and auditing time. 

\textbf{i) Initial Tree Creation Time.} Figure \ref{fig:perf}(a) shows our \texttt{EB-tree} outperforming other models, taking one-fourth the time of 8MHT and 75X less than blockchain models for tree creation.\\
\textbf{ii) Block Retrieval Time.} Figure \ref{fig:perf}(b) demonstrates our method's superior performance in block retrieval, taking less than 1 second for a 1GB file.\\
\textbf{iii) Block Update Time.} Our approach excels as file size increases, surpassing MHT and 8MHT in update time, shown in Figure \ref{fig:perf}(c).\\
\textbf{iv) Auditing Time.} Figure \ref{fig:perf}(d) illustrates our B-tree-based approach's significant improvement in auditing, consistently taking less than 1 second for a 1GB file, unlike other models that scale poorly with block count.

Unlike the conventional MHT and 8MHT-based auditing schemes (that are static in nature), our proposed approach supports dynamic batch auditing. Also, our approach uses a randomly generated seed value for every challenge, so that the file blocks can be retrieved every time to verify the challenge. This process eliminates the chance of false auditing and ensures the availability of data. 

\textbf{Dynamic Block Insertion and Deletion.} In contrast to alternative methods, our improved B-tree introduces support for block-level insertions and deletions. The time required for these operations as file sizes increase is illustrated in Table \ref{tab:ins-del}. The negligible amount of time taken for each operation suggests that this approach maintains efficiency even with larger file sizes.

\begin{table}[]
\resizebox{0.48\textwidth}{!}{%
\begin{tabular}{|l|r|r|r|r|r|r|}
\hline
\textbf{File Size} & \textbf{32MB} & \textbf{64MB} & \textbf{128MB} & \textbf{256MB} & \textbf{512MB} & \textbf{1GB} \\ \hline
\textbf{Total Blocks} & 21      & 41      & 82      & 164     & 328     & 656     \\ \hline
\textbf{Insert (sec)} & 0.0012 & 0.0019 & 0.0042 & 0.0099 & 0.0172 & 0.0362 \\ \hline
\textbf{Delete (sec)} & 0.0115 & 0.0329 & 0.0630 & 0.0996 & 0.1582 & 0.3188 \\ \hline
\end{tabular}%
}
\caption{Time required for insert and delete operations}
\label{tab:ins-del}
\end{table}

\section{Conclusion}
This paper presents a novel approach for dynamic auditing in centralized cloud environments using an enhanced B-tree data structure. Our scheme supports dynamic data operations while addressing performance challenges of decentralized architectures by constructing new tree nodes instead of overwriting existing ones. This approach ensures immutability and persistency for operations like update, insert, and delete. Cryptographic hashes and the use of private seed before hashing a block ensure data security and integrity. Our scheme is promising for secure cloud infrastructure supporting dynamic auditing, outperforming traditional MHT-based centralized auditing and decentralized blockchain-based auditing schemes in security, integrity, and immutability. Experimental results demonstrate superior time efficiency in terms of block modification, block retrieval, and batch auditing operations.

\bibliographystyle{IEEEtran}
\bibliography{IEEEabrv,references}

\end{document}